\documentclass[twocolumn,showpacs,preprintnumbers,amsmath,amssymb]{revtex4}
\usepackage{graphicx}% Include figure files
\usepackage{epsfig}
\usepackage{dcolumn}% Align table columns on decimal point
\usepackage{bm}% bold math
\def\bW {Br^{W{\to}l\nu}}
\def\bZ {Br^{Z{\to}ll}}
\def\os {opposite-sign }
\def\ktf {$k_t$-factorization }

\def\dps {DPS }
\def\chp#1#2#3  {{Chin.\ Phys. }             {\bf#1}, #2 (#3)}
\def\cpc#1#2#3  {{Computer\ Phys.\ Comm.\ }  {\bf#1}, #2 (#3)}
\def\err#1#2#3  {{\it Erratum }              {\bf#1}, #2 (#3)}
\def\epjc#1#2#3 {{Eur. Phys. J. C }          {\bf#1}, #2 (#3)}
\def\dum#1#2#3  {{~}                         {\bf#1}, #2 (#3)}
\def\ib#1#2#3   {{\it ibid. }                {\bf#1}, #2 (#3)}
\def\jcp#1#2#3  {{J.\ Comput.\ Phys.\ }      {\bf#1}, #2 (#3)}
\def\jetpl#1#2#3 {{\rm JETP Lett.}           {\bf#1}, #2 (#3)}
\def\jhep#1#2#3 {{JHEP }                     {\bf#1}, #2 (#3)}
\def\ijmp#1#2#3 {{Int.\ J.\ Mod.\ Phys.\ }   {\bf#1}, #2 (#3)}
\def\jpg#1#2#3  {{J.\ Phys.\ G }             {\bf#1}, #2 (#3)}
\def\mpl#1#2#3  {{Mod.\ Phys.\ Lett.\ }      {\bf#1}, #2 (#3)}
\def\ncim#1#2#3 {{Nuovo Cimento }            {\bf#1}, #2 (#3)}
\def\np#1#2#3   {{Nucl.\ Phys.\ }            {\bf#1}, #2 (#3)}
\def\npb#1#2#3   {{Nucl.\ Phys.\ B }         {\bf#1}, #2 (#3)}
\def\pan#1#2#3  {{Phys.\ At.\ Nuclei }       {\bf#1}, #2 (#3)}
\def\plb#1#2#3  {{Phys.\ Lett.\ B }          {\bf#1}, #2 (#3)}
\def\prep#1#2#3 {{Phys.\ Rep.\ }             {\bf#1}, #2 (#3)}
\def\prd#1#2#3  {{Phys.\ Rev.\ D }           {\bf#1}, #2 (#3)}
\def\prl#1#2#3  {{Phys.\ Rev.\ Lett.\ }      {\bf#1}, #2 (#3)}
\def\ptp#1#2#3  {{Prog.\ Theor.\ Phys.\ }    {\bf#1}, #2 (#3)}
\def\ps#1#2#3   {{Physica Scripta }          {\bf#1}, #2 (#3)}
\def\rmp#1#2#3  {{Rev.\ Mod.\ Phys.\ }       {\bf#1}, #2 (#3)}
\def\rpp#1#2#3  {{Rep.\ Prog.\ Phys.\ }      {\bf#1}, #2 (#3)}
\def\sa#1#2#3   {{Sci. Acta}                 {\bf#1}, #2 (#3)}
\def\sjnp#1#2#3 {{Sov.\ J.\ Nucl.\ Phys.\ }  {\bf#1}, #2 (#3)}
\def\spj#1#2#3  {{Sov.\ Phys.\ JETP }        {\bf#1}, #2 (#3)}
\def\spjl#1#2#3 {{Sov.\ JETP Lett.\ }        {\bf#1}, #2 (#3)}
\def\spu#1#2#3  {{Sov.\ Phys.-Usp.\ }        {\bf#1}, #2 (#3)}
\def\yaf#1#2#3  {{Yad.\ Fiz.\ }              {\bf#1}, #2 (#3)}
\def\zp#1#2#3   {{Zeit.\ Phys.\ }            {\bf#1}, #2 (#3)}
\def\zpc#1#2#3  {{Z.\ Phys.\ C }             {\bf#1}, #2 (#3)}
\def\et {{\it et al. }}
\begin{document}
%\title{Associated $WD$ and $WB$ production at LHCb\\
%     and prospects to observe double parton interactions}
\title{Associated production of electroweak bosons and heavy mesons at LHCb\\
     and the prospects to observe double parton interactions}
\author{S.\ P.\ Baranov}
\email{baranov@sci.lebedev.ru}
\affiliation{P.N. Lebedev Institute of Physics, 
             53 Lenin Avenue, Moscow 119991, Russia}
\author{A.\ V.\ Lipatov}
\email{lipatov@theory.sinp.msu.ru}
\affiliation{Skobeltsyn Institute of Nuclear Physics,
     Lomonosov Moscow State University, Moscow 119991, Russia}
\affiliation{Joint Institute for Nuclear Research, Dubna 141980, Moscow
region, Russia}
\author{M.\ A.\ Malyshev}
\email{malyshev@theory.sinp.msu.ru}
\affiliation{Skobeltsyn Institute of Nuclear Physics,
     Lomonosov Moscow State University, Moscow 119991, Russia}
\author{A.\ M.\ Snigirev}
\email{snigirev@lav01.sinp.msu.ru}
\affiliation{Skobeltsyn Institute of Nuclear Physics,
     Lomonosov Moscow State University, Moscow 119991, Russia}
\author{N.\ P.\ Zotov\footnote{Deceased}}
%\email{zotov@theory.sinp.msu.ru}
\affiliation{Skobeltsyn Institute of Nuclear Physics,
      Lomonosov Moscow State University, Moscow 119991, Russia}
\date{\today}
\begin{abstract}
The production of weak gauge bosons in association with heavy flavored mesons 
at the LHCb conditions is considered, and a detailed study of the different
contributing processes is presented including single and double (DPS) 
parton scattering mechanisms. 
We find that the usual DPS factorization formula 
needs to be corrected for the limited partonic phase space, and that including 
the relevant corrections reduces discrepancies in the associated $ZD$ production. 
We conclude finally that double parton scattering dominates the production of 
same-sign $W^\pm D^\pm$ states, as well as the production of $W^-$ bosons 
associated with $B$-mesons.
The latter processes can thus be regarded as new useful DPS indicators.

\end{abstract} 
\maketitle

\section{Introduction}

In our recent publication \cite{WD1} we have considered the associated production
of charged gauge bosons $W^\pm$ and charged charmed mesons $D^{(*)\pm}$ at the 
LHC and came to the conclusion that same-sign $W^\pm D^{(*)\pm}$ events could 
serve as an indicator of double parton interactions 
\cite{Bartalini,Abramowicz,Bansal}. Our consideration was only restricted to the 
central region, i.e. to CMS \cite{CMS} and ATLAS \cite{ATLAS} kinematic 
conditions, since these were the only collaborations who provided the data 
(though not on same-sign $WD$ configurations).
%Soon after the acceptance of our paper, we have learned \cite{LHCb} that 
To the best of our knowledge, the LHCb 
Collaboration is going to measure the production cross sections for all of the 
four $WD$ charge combinations. Now, we feel it very tempting to foreshadow the
experimental measurement with a theoretical prediction. 

The planned work needs to be done with care, since the momentum conservation 
requirement in the large-$x$ region may spoil the factorization hypothesis 
commonly used in double parton scattering (DPS) calculations. This motivates
us to introduce certain corrections to the theory. We also wish 
to extend our analysis to the production of $WB$ states. The latter is closely 
similar to the $WD$ case in its \dps part, but the 
background coming from single parton scattering (SPS) is rather different 
and awaits a special survey.

Our previous calculation \cite{WD1} was done in the $k_t$-factorization 
technique, but in the present paper we adopt `combined' approach. That is,
the production of heavy systems like $W$ or $Z$ bosons as well as their SPS
production in association with heavy quarks is done in the traditional
collinear scheme, while the $k_t$-factorization is used for solely produced 
$c\bar{c}$ or $b\bar{b}$ pairs (the latter constitute one branch of a double
parton interaction). Then we benefit from easily including higher-order 
radiative corrections which are taken into account in the form of $k_t$-%
dependent parton densities.
Thus, we rely on a combination of two techniques, with each of them being 
used at the kinematics where it is most suitable ($W$ and $Z$ at large $x$, 
$c\bar{c}$ and $b\bar{b}$ at small $x$).

The outline of the paper is as follows. In Sec. II, we reconsider the DPS
formalism in the forward (LHCb) kinematics and introduce corrections matching
the momentum conservation requirement. Then we test our theory by applying 
it to the associated $ZD$ production, where the existing data \cite{LHCb-ZD} 
form grounds for a comparison.
We further use the corrected formalism to make predictions on the charm-%
associated $W^\pm$ production in Sec. III and on the beauty-associated $W^\pm$
production in Sec. IV. Our findings are summarized in Sec. V.

\section{Double parton scattering in the large-$x$ region}

As far as the SPS contributions are concerned, the calculations are
straightforward and need no special explanation.
%Examples of the relevant Feynman diagrams are shown in Fig. 1.
Throughout this paper, all calculations are based on the following parameter 
setting.
We employ the \ktf approach \cite{GLR,smallX} for relatively light states 
($c\bar{c}$ or $b\bar{b}$) and collinear factorization for states containing 
$W$ or $Z$ bosons.
We used Kimner-Martin-Ryskin \cite{KMR} parametrization for unintegrated quark and gluon 
distributions with Martin-Stirling-Thorne-Watt (MSTW) \cite{MSTW} collinear densities taken as input
(or pure MSTW densities for collinear calculations);
we used running strong and electroweak coupling constants normalized to
$\alpha_s(m^2_Z){=}0.118$; $\alpha(m^2_Z){=}1/128$; $\sin^2\Theta_W{=}0.2312$;
the factorization and renormalization scales were chosen as
$\mu^2_R$=$\mu^2_F$=$m^2_T(W)$ $\equiv$ $m^2_W{+}p^2_T(W)$ or $m^2_T(Z)$ for
all processes involving $W$ and $Z$ bosons, and  $\mu^2_R$=$\mu^2_F$=$m^2_Q$
%$\mu^2_R$=$\mu^2_F$=$(m^2_T(Q){+}m^2_T(\bar{Q}))/2$ 
for the production of sole $Q\bar{Q}$ pairs ($Q=c,b$);
the quark masses were set to $m_c{=}1.5$~GeV, $m_b{=}4.5$~GeV, $m_t{=}175$~GeV,
$c$- and $b$-quarks were converted into $D^+$ and $B$ mesons
using the Peterson fragmentation function \cite{Peterson}
with $\epsilon_c{=}0.06$ and $\epsilon_b{=}0.006$, respectively,
and normalized to $f(c{\to}D^+)=0.268$ \cite{JKLZ},
$f(b{\to}B^-)$=0.40 and $f(b{\to}\bar{B}^0)$=0.40.

Our choice of renormalization scale is slightly different from the
conventional one by using $m_Q$ rather than $m_T(Q)$, but we then can
fit the experimental data (see below, eqs. (4),(15),(22); otherwise,
with $\mu^2_R$=$m^2_T(Q)\equiv m^2_Q{+}p^2_T(Q)$, the calculations
would lie slightly below the data points). We do not mind developing
here a rigorous theory of heavy quark production, but are rather
interested in understanding the relative importance of the different
contributions. Our simple prescription would suffice for that purpose.

To calculate the DPS contributions one commonly makes use of a simple factorization 
formula (for details see the reviews \cite{Bartalini,Abramowicz,Bansal} and 
references therein),
\begin{equation} \label{doubleAB}
\sigma^{WD}_{\rm DPS} =
\sigma^{W}_{\rm SPS}\sigma^{D}_{\rm SPS}/\sigma_{\rm eff},
\end{equation}
where $\sigma_{\rm eff}$ is a normalization constant that encodes all 
``DPS unknowns'' into a single phenomenological parameter.  
%A numerical value of $\sigma_{\rm eff}\simeq$ 15 mb has earlier been obtained 
%empirically from fits to $p\bar{p}$ and $pp$ data.
Deriving this formula relies on two simplifying approximations: that
(i) the double parton distribution functions can be decomposed into longitudinal 
and transverse components, and 
(ii) the longitudinal component $D^{ij}_p(x_1,x_2; Q^2_1,Q^2_2)$  reduces to the 
diagonal product of two independent single parton distribution functions:
\begin{eqnarray}\label{DxD}
D^{ij}_p(x_1, x_2; Q^2_1, Q^2_2) = D^i_p (x_1; Q^2_1) D^j_p (x_2; Q^2_2)
\end{eqnarray}
(here $x_1$ and $x_2$ are the longitudinal momentum fractions of the partons $i$ 
and $j$ entering the hard subprocesses at the probing scales $Q_1$ and $Q_2$). 
The latter approximation is acceptable for such collider experiments where only 
small $x$ values are probed; however, this cannot be said of the LHCb conditions,
especially with respect to heavy systems as electroweak bosons.
At the LHCb conditions, the probed $x$ values are not far from the phase 
space boundary where the evident restriction on the total parton momentum
$x_1{+}x_2\leq 1$ violates the DPS factorization ansatz.

Setting 
the boundary condition in the form of theta-function $\Theta(1{-}x_1{-}x_2)$ 
would result in a steplike discontinuity at the edge of the phase space. 
This does not seem physically consistent for the parton densities.
In a more accurate approach \cite{snig04,Gaunt:2009re,Snigirev:2010ds,Chang:2012nw,%
Rinaldi:2013vpa,Golec-Biernat:2014bva,Ceccopieri:2014ufa,Snigirev:2014eua},
\begin{eqnarray}\label{DxDr}
D^{ij}_p(x_1, x_2; Q^2_1, Q^2_2) = D^i_p (x_1; Q^2_1) D^j_p (x_2; 
Q^2_2)\nonumber\\
\times (1{-}x_1{-}x_2)^n,
\end{eqnarray}
the kinematical constraints are smoothly put into play with the 
correction factor $(1{-}x_1{-}x_2)^n$, where $n>0$ is a parameter to be fixed 
phenomenologically. The integrand and its derivative remain continuous at
the phase space border. One often chooses n=2. This choice of the 
phase space factor can be partly justified~\cite{snig04,Snigirev:2010ds} 
in the framework of perturbative QCD and gives double parton distribution 
functions which satisfy the momentum sum rules~\cite{Gaunt:2009re} reasonably 
well. To feel the size of the possible effect we also tried $n=3$. 
The case of unconstrained phase space is presented in Table~I as $n=0$.
A numerical value of $\sigma_{\rm eff}\simeq$ 15 mb has earlier been
obtained empirically from fits to $p\bar{p}$ and $pp$ data. This will be
taken as the default value throughout the paper. As we  will see,
variations within some reasonable range $\sigma_{\rm eff}\simeq$ 15
$\pm$ 5 mb would affect our DPS predictions (with the respective errors
presented in the tables), though without changing our basic conclusions.

Now we are ready to compare predictions with the data.
For the LHCb fiducial phase space \cite{LHCb-ZD} we obtain
\begin{eqnarray}
\sigma_{\mbox{incl}}(D^+)+\sigma_{\mbox{incl}}(D^0)&{=}& 670\;\mu\mbox{b},\\ 
\bZ\sigma_{\mbox{incl}}(Z^0)&{=}&75\;\mbox{pb},\label{z}
\end{eqnarray}
in excellent agreement with Ref. \cite{LHCb-Z}, reporting  
$\bZ\sigma_{\mbox{incl}}(Z^0)=76\;\mbox{pb}$.

As the experimental statistics is very limited (7 $ZD^0$ events and 4 $ZD^+$ 
events) it is more reasonable not to consider the $ZD^0$ and $ZD^+$ cross 
sections separately, but rather to rely on the sum of them. {%\footnote{%
Taken separately, the $ZD^0$ and $ZD^+$ data are at variance with other
measurements. There exist independent publications \cite{LHCb-D,LEP-D,JKLZ}
(including the one by LHCb Collaboration) which all agree with each other
showing the ratio $\sigma(D^0)/\sigma(D^+)\sim 2.5$, in contrast with
$\sigma(ZD^0)/\sigma(ZD^+)\sim 5.5$ seen in \cite{LHCb-ZD}. In fact, the 
authors of \cite{LHCb-ZD} seem to greatly underestimate their statistical 
errors.} 

So, we calculate the $Zc\bar{c}$ production cross section at the quark level 
and then convert $c$-quarks into $D^0$ and $D^+$ mesons with the overall probability 
normalised to 85\% (with the remaining 15\% left for $D_s$ and $\Lambda_c$).
We estimate the yields from the different subprocesses as
\begin{eqnarray}
\sigma(u\bar{u}\to Zc\bar{c}) &=& 5\;\mbox{pb},\\
\sigma(d\bar{d}\to Zc\bar{c}) &=& 2.6\;\mbox{pb},\\
\sigma(gu\to Zuc\bar{c}) &=& 11.4\;\mbox{pb},\\
\sigma(gd\to Zdc\bar{c}) &=& 5.2\;\mbox{pb},\\
\sigma(gg\to Zc\bar{c}) &=& 2.5\;\mbox{pb}.
\end{eqnarray}
Summing up and multiplying by the quark fragmentation probability and
by the $Z\to\mu^+\mu^-$ branching fraction we arrive at 
$\sigma^{SPS}(ZD^0,ZD^+)=0.85$ pb. This result is consistent with the theoretical
calculation presented in \cite{LHCb-ZD} under the name of `MCFM massive'.
Adding the DPS contribution in the form (\ref{doubleAB}) gives
$\sigma^{SPS+DPS}(ZD^0,ZD^+)=4.2$ pb, which significantly exceeds the data.
After applying the correction factor (\ref{DxDr}) the agreement becomes
rather satisfactory (see Table I).

\begin{table}[t]\label{tabZD}
\caption{Comparison of the measured and predicted cross-sections (in pb) for 
$Z$ bosons produced in association with open charm mesons in the fiducial region
$p_T(\mu^\pm)>20$ GeV, $2<\eta(\mu^\pm)<4.5$, $2<p_T(D)<12$ GeV, $2<y(D)<4$.
The SPS and DPS contributions are shown separately, with $n$ indicating the power
of the correction factor in Eq.(\ref{DxDr})}
\begin{tabular}{lccccc}
\hline
channel    &~data~&~~~SPS~~~&~DPS(n{=}0)&~DPS(n{=}2)&~DPS(n{=}3)\\
%$Z^0D^0$  & 2.50 &   0.6   &   3.3     &   1.6      &    1.2   \\
%$Z^0D^+$  & 0.44 &   0.25  &   1.3     &   0.6      &    0.5   \\
%sum       & 2.94 &   0.85  &   4.6     &   2.2      &    1.7   \\
$Z^0D^0$   & 2.50 &   0.6   &   $2.4\pm0.6$     &   $1.15\pm0.38$     &    $0.95\pm0.32$  \\
$Z^0D^+$   & 0.44 &   0.25  &   $0.95\pm0.32$    &   $0.50\pm0.17$      &    $0.40\pm0.13$   \\
sum        & 2.94 &   0.85  &   $3.35\pm0.92$    &   $1.65\pm0.55$     &    $1.35\pm0.45$  \\
\hline
\end{tabular}
\end{table}

\section{Charm-associated $W^\pm$ production}

%As far as the single parton scattering is concerned, we use exactly the same 
%model and parameter setting as we did previously in Ref. \cite{WD1}.
%We only remind here some essential points for the reader's convenience. 

The production of \os $W^{\pm}D^{\mp}$ states is dominated by the 
quark-gluon scattering at ${\cal O}(\alpha_s\alpha)$
\begin{equation}\label{gq2wc}
g+q\to W^-+c \mbox{~~~~or~~~~} g+\bar{q}\to W^++\bar{c},
\end{equation}
where the main role belongs to strange quarks, $q{=}s$.
Among the variety of processes contributing to both \os and same-sign $WD$ states,
the most important ones are the quark-antiquark annihilation at 
${\cal O}(\alpha_s^2\alpha)$,
\begin{equation}\label{qq2wcc}
u+\bar{d}\to W^+ + c + \bar{c} \mbox{~~or~~} d+\bar{u}\to W^- + c + \bar{c},
\end{equation}
and quark-gluon scattering at ${\cal O}(\alpha_s^3\alpha)$,
\begin{equation}\label{gq2qwcc}
g + u\to W^+ + d + c + \bar{c} \mbox{~~or~~} g + d\to W^- + u + c + \bar{c}.
\end{equation}
In addition to that, there present indirect contributions from the production
of top-quark pairs
\begin{equation}\label{gg2tt}
g+g\to t+\bar{t} \mbox{~~and~~} q+\bar{q}\to t+\bar{t}
\end{equation}
followed by a long chain of decays:
$t\to W^+b$, $W^+\to c\bar{s}$, $b\to cX$ or $b\to c\bar{c}s$ 
(and the charge conjugated modes). 
All other possible processes are suppressed by extra powers of coupling 
constants or by Kobayashi-Maskawa mixing matrix.
Subprocesses $q\bar{q}\to W^-c\bar{s}$ and $q\bar{q}\to W^+s\bar{c}$, though 
formally of the same order as (\ref{qq2wcc}), are heavily suppressed by the 
gluon propagator having vitrtuality of order $m_W^2$ rather than $m_{cc}^2$.

Our parameter setting was basically described in Sec. II.
For the indirect contributions we also assumed $100\%$ branching fraction 
for $t\to bW$ 
%equal fragmentation probabilities for $b\to\bar{B}^0$ and $b\to B^-$,
and used the inclusive branching fractions 
$Br(\bar{B}^0{\to}D^+X)=37\%$, $Br(B^0{\to}D^+X)=3\%$,
$Br(B^-{\to}D^+X)=10\%$ and $Br(B^+{\to}D^+X)=2.5\%$
listed in the Particle Data Book \cite{PDG}.

The evaluation of the \dps contributions is done in
accordance with the explanations given in the previous section.
%A numerical value of $\sigma_{\rm eff}\simeq$ 15 mb has earlier been obtained
%empirically from fits to $p\bar{p}$ and $pp$ data. 
The individual inclusive SPS cross sections $\sigma(D^{\pm})$ and $\sigma(W^{\pm})$ 
have been calculated as in Refs. \cite{JKLZ} and \cite{inclW}, respectively. 
For the LHCb fiducial phase space our expectations read
\begin{eqnarray}
\sigma_{\mbox{incl}}(D^+)=\sigma_{\mbox{incl}}(D^-)&{=}& 190\;\mu\mbox{b},\\
\bW\sigma_{\mbox{incl}}(W^+)&{=}&970\;\mbox{pb},\label{wp}\\
\bW\sigma_{\mbox{incl}}(W^-)&{=}&680\;\mbox{pb},\label{wm}
\end{eqnarray}
in good agreement with \cite{LHCb-ZD} and \cite{LHCb-W}, respectively.
Our results for SPS and DPS channels are displayed in Table II.
All DPS contributions are presented there without phase space corrections; they
have to be multiplied by a correction factor of 0.48 for n=2 or 0.38 for n=3.

\begin{table}[t]\label{tabWD}
\caption{Predicted $WD$ production cross sections times the $W{\to}l\nu$ 
branching (in pb) integrated over the fiducial region 
$p_T(l)>20$ GeV, $2<\eta(l)<4.5$, $2<p_T(D)<12$ GeV, $2<\eta(D)<4$}

\begin{tabular}{lcccc}
%subprocess  &$\;\;W^+D^+\;$ &$\;\;W^+D^-\;$ &$\;\;W^-D^-\;$ &$\;\;W^-D^+\;$\\
\hline
\multicolumn{5}{c}{Double parton scattering contributions}\\
subprocess  &$\;\;W^+D^+\;$ &$\;\;W^+D^-\;$ &$\;\;W^-D^-\;$ &$\;\;W^-D^+\;$\\
%\hline
$gg{\to}c\bar{c},\;\;u\bar{d}{\to}W^+$ & $12.3\pm4.1$ & $12.3\pm4.1$  &  --   &  --   \\
$gg{\to}c\bar{c},\;\;d\bar{u}{\to}W^-$ &  --  &   --  &  $8.9\pm3.0$  & $8.9\pm3.0$   \\
%                                    ~ &  ~   &   ~   &  ~    &  ~    \\
\hline
\multicolumn{5}{c}{Single parton scattering contributions}\\
subprocess  &$\;\;W^+D^+\;$ &$\;\;W^+D^-\;$ &$\;\;W^-D^-\;$ &$\;\;W^-D^+\;$\\
%\hline
$g\bar{s},g\bar{d}{\to}W\bar{c}$   &   --   &  1.7   &  --    &  --    \\
$gs,gd{\to}Wc$                     &   --   &  --    &  --    &  2.0   \\
$u\bar{d}{\to}Wc\bar{c}$           &  0.8   &  0.8   &  --    &  --    \\
$d\bar{u}{\to}Wc\bar{c}$           &   --   &   --   &  0.4   &  0.4   \\
$gu{\to}Wdc\bar{c}$                &  1.9   &  1.9   &  --    &  --    \\
$g\bar{d}{\to}W\bar{u}c\bar{c}$    &  0.16  &  0.16  &  --    &  --    \\
$gd{\to}Wuc\bar{c}$                &   --   &  --    &  0.8   &  0.8   \\
$g\bar{u}{\to}W\bar{d}c\bar{c}$    &   --   &  --    &  0.14  &  0.14  \\ 
$gg{\to}t\bar{t}{\to}$decays       &  0.01  &  0.01  &  0.01  &  0.01  \\
$q\bar{q}{\to}t\bar{t}{\to}$decays &  0.015 &  0.02  &  0.015 &  0.02  \\
\hline
\end{tabular}
\end{table}

The indirect contributions, though small already, can be further suppressed
using a well-known experimental technique based on the
property that the secondary $b$-decay vertex is displaced with respect to
the primary interaction vertex. Summing up the direct contributions, we see
that the predicted same-sign $WD$ production rates with and without DPS channels 
differ by a significant factor. This difference is sensible enough to 
%provide reliable grounds for interpreting 
warrant interpretation of
the forthcoming LHCb data as giving conclusive evidence for double parton 
interactions.

\section{Beauty-associated $W^\pm$ production}

Associated $WB$ production is not simply a repetition of the $WD$ case with
a different quark mass. Indeed, the contributing parton subprocesses are
significantly different. First, there is no analog to process (\ref{gq2wc}), 
as the Cabibbo-Kobayashi-Maskawa couplings of a $b$-quark to the quarks of two lighter generations 
are really negligible.
Second, the feed down from top-quark decays now must be regarded as a direct 
contribution, as it shows no secondary decay vertex (and, therefore, cannot
be rejected experimentally).

The full list of processes included in the present analysis reads as follows:
quark-antiquark annihilation at ${\cal O}(\alpha_s^2\alpha)$,
\begin{equation}\label{qq2wbb}
u+\bar{d}\to W^+ + b + \bar{b} \mbox{~~or~~} d+\bar{u}\to W^- + b + \bar{b};
\end{equation}
quark-gluon scattering at ${\cal O}(\alpha_s^3\alpha)$,
\begin{equation}\label{gq2qwbb}
g + u\to W^+ + d + b + \bar{b} \mbox{~~or~~} g + d\to W^- + u + b + \bar{b};
\end{equation}
strong production of top-quark pairs
\begin{equation}\label{gg2bb}
g+g\to t+\bar{t} \mbox{~~and~~} q+\bar{q}\to t+\bar{t}
\end{equation}
followed by their decays $t{\to}W^+b$, $\bar{t}{\to}W^-\bar{b}$;
and, finally, weak production of $t\bar{b}$ and $\bar{t}b$ states
\begin{equation}\label{qq2tb}
u+\bar{d}\to t+\bar{b} \mbox{~~or~~} d+\bar{u}\to b+\bar{t},
\end{equation}
also followed by $t$-decays.

With the parameter setting described in Sec. III, we estimate the 
inclusive production of $b$ quarks in the LHCb domain as
\begin{equation}\label{bb}
\sigma_{\mbox{incl}}(b)=\sigma_{\mbox{incl}}(\bar{b}) = 95\;\mu\mbox{b}.
\end{equation}
This number is compatible with the experimental result \cite{LHCb-B}
\begin{equation}
%\sigma(B^+)+\sigma(B^0)+\sigma(B_s) = 39\;\mu{b} + 38\;\mu{b} + 10\;\mu{b}
%                                    = 87\;\mu{b};
\sigma(B^+)+\sigma(B^0)+\sigma(B_s) = 39 + 38 + 10 = 87\;\mu{b};
\end{equation}
at least, it lies within the usual theoretical uncertainty related to 
the choice of the interaction scale and quark mass. Combining this result 
with Eqs.(\ref{wp}), (\ref{wm}) we obtain the DPS cross section for $Wb$.
Table III represents our predictions for unconstrained phase space of
Eq. (\ref{doubleAB}); they have to be corrected by a factor of 0.45 for 
n=2 or 0.36 for n=3.

We find it worth saying a few words on 
%instructive to shed more light on 
the specific properties of SPS and DPS kinematics at the LHCb conditions.
Parton momentum configurations in the SPS channels are very asymmetric.
To produce a heavy $Wb$ system with both $W$ and $b$ having large positive
rapidity, the positive light-cone momentum fraction of the incoming parton
must be large. On the average, valence $u$ quarks carry larger $x$ than 
valence $d$ quarks, thus favoring the production of $W^+$ in comparison 
with $W^-$ bosons in subprocesses (\ref{qq2wbb}) and (\ref{gq2qwbb}).
This property is illustrated in Figs. 1, 2. 
Sea quarks are mainly concentrated in the small-$x$ region and cannot 
contribute at a significant level.

In general, the desired large positive light-cone momentum is easier to
get with two independent partons in DPS than with a single parton in SPS.
This explains the relative suppression of the SPS channels seen in Table III. 
Especially pleasant are negligible contributions from top-quark decays.
Their rapidity distributions are shown in Fig. 3. DPS clearly and
unambiguously dominates the production of $W^\pm B$ states, making them very
informative observables.

\begin{table}[t]\label{tabWB}
\caption{Predicted $WB$ production cross sections times the $W{\to}l\nu$
branching (in pb) integrated over the fiducial region
$p_T(l)>20$ GeV, $2<\eta(l)<4.5$, $2<\eta(B)<4.5$. 
Here $B^+$ and $B^-$ denote the sum of $B^+$ and $B^0$ and the sum of 
$B^-$ and $\bar{B}^0$ mesons, respectively.}

\begin{tabular}{lcccc}
%subprocess  &$\;\;W^+B^+\;$ &$\;\;W^+B^-\;$ &$\;\;W^-B^-\;$ &$\;\;W^-B^+\;$\\
\hline
\multicolumn{5}{c}{Double parton scattering contributions}\\
subprocess  &$\;\;W^+B^+\;$ &$\;\;W^+B^-\;$ &$\;\;W^-B^-\;$ &$\;\;W^-B^+\;$\\
%\hline
$gg{\to}b\bar{b},\;\;u\bar{d}{\to}W^+$ & $5.5\pm1.8$  &  $5.5\pm1.8$  &  --   &  --   \\
$gg{\to}b\bar{b},\;\;d\bar{u}{\to}W^-$ &  --  &   --  &  $4.0\pm1.3$  &  $4.0\pm1.3$  \\    
\hline
%                                    ~ &  ~   &   ~   &  ~    &  ~    \\
\multicolumn{5}{c}{Single parton scattering contributions}\\
subprocess  &$\;\;W^+B^+\;$ &$\;\;W^+B^-\;$ &$\;\;W^-B^-\;$ &$\;\;W^-B^+\;$\\
%\hline
$u\bar{d}{\to}Wb\bar{b}$           &  1.2  &  1.2  &  --   &  --   \\
$d\bar{u}{\to}Wb\bar{b}$           &   --  &   --  &  0.5  &  0.5  \\
$gu{\to}Wdb\bar{b}$                &  2.7  &  2.7  &  --   &  --   \\
$g\bar{d}{\to}W\bar{u}b\bar{b}$    &  0.22 &  0.22 &  --   &  --   \\
$gd{\to}Wub\bar{b}$                &   --  &  --   &  1.1  &  1.1  \\
$g\bar{u}{\to}W\bar{d}b\bar{b}$    &   --  &  --   &  0.2  &  0.2  \\
$gg{\to}t\bar{t}{\to}WWb\bar{b}$       & 0.030 & 0.045 & 0.030 & 0.045  \\
$q\bar{q}{\to}t\bar{t}{\to}WWb\bar{b}$ & 0.055 & 0.060 & 0.055 & 0.060  \\
$u\bar{d}{\to}t\bar{b}{\to}Wb\bar{b}$  & 0.0018& 0.0042& 0.0018& 0.0042 \\
$d\bar{u}{\to}b\bar{t}{\to}W\bar{b}b$  & 0.0002& 0.0005& 0.0002& 0.0005 \\
\hline
\end{tabular}
\end{table}

\section{Conclusions}

Having considered the production of $Z^0D$, $W^\pm D$ and $W^\pm B$ 
states at the LHCb conditions we deduce the following assessments.

(i) As a general rule for the production of electroweak bosons in the DPS
channel, the simple DPS factorization formula needs to be corrected for
the limited partonic phase space. Numerically, these corrections amount
to a factor of 2 in the total rates and, when taken into account, lead to
better agreement with the available data on $Z^0D$ production than there 
seemed to be before.

(ii) Similarly to what has been observed earlier for the central region 
(ATLAS and CMS), the production of same-sign $W^\pm D^\pm$ states in the
forward region is also dominated by the DPS mechanism. Once again, this
process can be recommended as a DPS indicator.

(iii) Along with that, LHCb kinematics opens doors for a still new 
indicative process, which is the beauty-associated production of weak
gauge bosons $W^-$. The charge of the accompanying $b$ quark is irrelevant.
Here we benefit from the asymmetric rapidity selection cuts, which correspond
to large positive light-cone momentum values of the incoming partons.
The essential values can easier be reached with two independent partons in 
DPS than with a single parton in SPS, thus giving favor to DPS production.
Another useful feature of the LHCb kinematics as compared to ATLAS and CMS
conditions is in much lower $p_t$ cuts for inclusive open flavor production. 
This enhances the visible inclusive cross sections $\sigma_{\mbox{incl}}(D)$ 
and $\sigma_{\mbox{incl}}(B)$ and, consequently, the DPS channel in associated 
production with gauge bosons.

\section*{Acknowledgments}

The authors would like to thank I. Belyaev for the useful discussions.
This work was supported in part by RFBR grant 16-32-00176-mol-a, President of Russian Federation Grant NS-7989.2016.2, and by the DESY Directorate in the framework of the Moscow-DESY project
on Monte Carlo implementations for HERA-LHC.

%\end{document}

\vspace*{1cm}
%\caption{Examples of Feynman diagrams representing the relevant partonic
%subprocesses. Solid lines, quarks; twirly lines, gluons; zigzag lines,
%electroweak bosons.}

\begin{figure}\label{w-}
\epsfig{figure=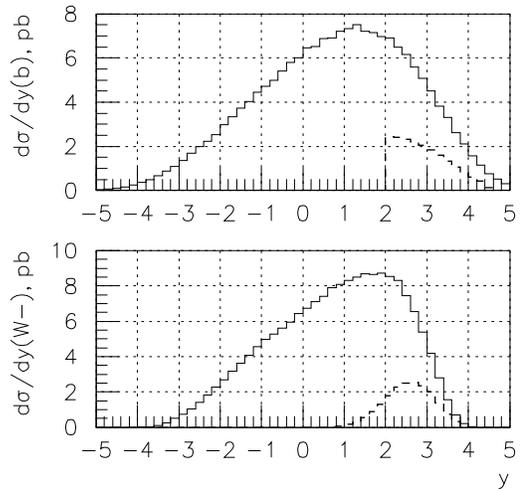,width=8.5cm}
\caption{Rapidity distributions of the $b$ quarks (upper panel) and
$W^-$ bosons (lower panel) produced in association in the process
$d\bar{u}{\to}W^-b\bar{b}$. Solid curves, original spectra; dashed curves,
left after imposing the LHCb kinematic cuts.}
\end{figure}

\begin{figure}\label{w+}
\epsfig{figure=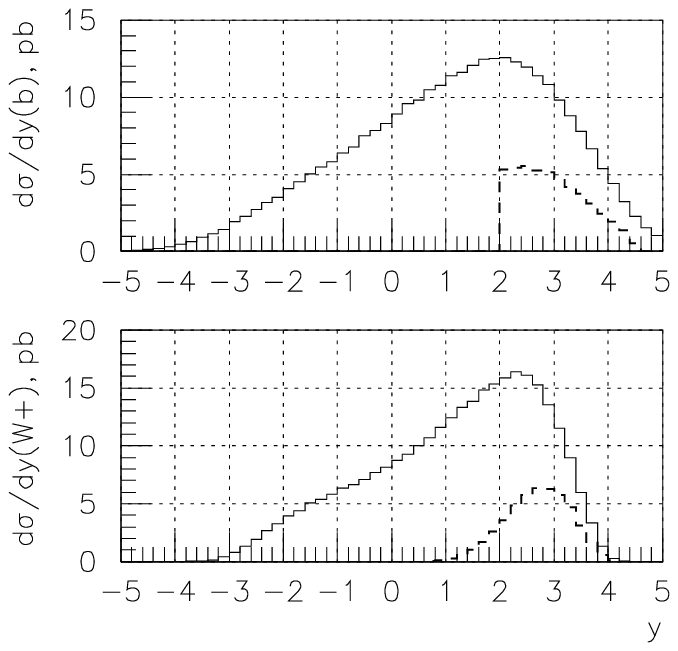,width=8.5cm}
\caption{Rapidity distributions of the $b$ quarks (upper panel) and
$W^+$ bosons (lower panel) produced in association in the process
$u\bar{d}{\to}W^+b\bar{b}$. Solid curves, original spectra; dashed curves,
left after imposing the LHCb kinematic cuts.}
\end{figure}

\begin{figure}\label{yt}
\epsfig{figure=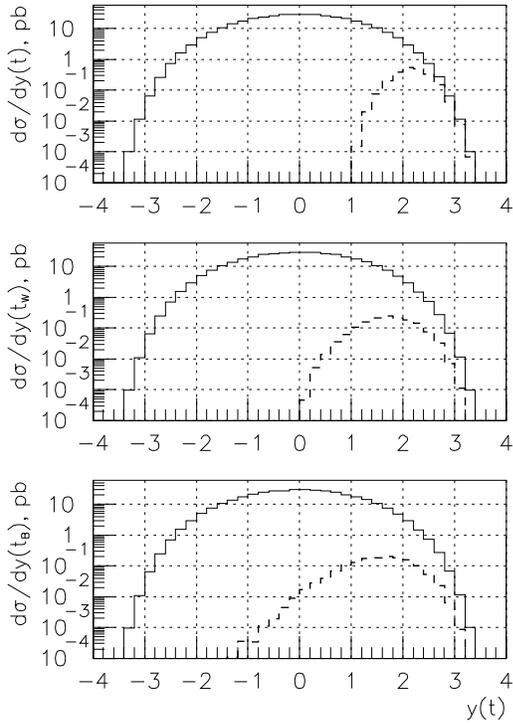,width=8.5cm}
\caption{Rapidity distributions of the top-quarks or antiquarks.
Opposite sign $W^\pm b^\mp$ events: top quarks converting into a $W^\pm b^\mp$ 
pair (upper panel).
Same sign $W^\pm b^\pm$ events: top quarks producing W bosons (middle panel);
top quarks producing beauty quarks (lower panel).
Solid curves, original spectra;  dashed curves, left after imposing the LHCb 
kinematic cuts.}
\end{figure}

\end{document}